# Anomalous continuum scattering and higher-order van Hove singularity in the strongly anisotropic *S* = 1/2 triangular lattice antiferromagnet

Pyeongjae Park[1*], E. A. Ghioldi[2], Andrew F. May[1], James A. Kolopus[1], Andrey A. Podlesnyak[3], Stuart Calder[3], Joseph A. M. Paddison[1], A. E. Trumper[4], L. O. Manuel[4], Cristian D. Batista[2,5], Matthew B. Stone[3*], Gábor B. Halász[1*] & Andrew D. Christianson[1*]

[1]*Materials Science and Technology Division, Oak Ridge National Laboratory, Oak Ridge, Tennessee 37831, USA*

[2]*Department of Physics and Astronomy, University of Tennessee, Knoxville, Tennessee 37996, USA*

[3]*Neutron Scattering Division, Oak Ridge National Laboratory, Oak Ridge, Tennessee 37831, USA*

[4]*Instituto de Física Rosario (CONICET) and Universidad Nacional de Rosario, Rosario, Argentina*

[5]*Neutron Scattering Division and Shull-Wollan Center, Oak Ridge National Laboratory; Oak Ridge, Tennessee 37831, USA*

*Corresponding author: parkp@ornl.gov, stonemb@ornl.gov, halaszg@ornl.gov, and christiansad@ornl.gov

## Abstract

The $S = 1/2$ triangular lattice antiferromagnet (TLAF) is a paradigmatic example of frustrated quantum magnetism. An ongoing challenge involves understanding the influence of exchange anisotropy on the collective behavior within such systems. Using inelastic neutron scattering (INS) and advanced calculation techniques, we have studied the low and high-temperature spin dynamics of $Ba_2La_2CoTe_2O_{12}$ (BLCTO): a $Co^{2+}$-based $J_{\text{eff}} = 1/2$ TLAF that exhibits 120° order below $T_N = 3.26$ K. We determined the spin Hamiltonian by fitting the energy-resolved paramagnetic excitations measured at $T > T_N$, revealing exceptionally strong easy-plane XXZ anisotropy. Below $T_N$, the excitation spectrum exhibits a high energy continuum having a larger spectral weight than the single-magnon modes, suggesting a scenario characterized by a spinon confinement length that markedly exceeds the lattice spacing. We conjecture that this phenomenon arises from the proximity to a quantum melting point, even under strong easy-plane XXZ anisotropy. Finally, we highlight characteristic flat features in the excitation spectrum, which are connected to higher-order van Hove singularities in the magnon dispersion directly induced by easy-plane XXZ anisotropy. Our results provide a rare experimental insight into the nature of highly anisotropic $S = 1/2$ TLAFs between the Heisenberg and XY limits.

**Introduction**

Magnetic frustration plays an indispensable role in the formation of exotic states of matter. At the level of nearest-neighbor interactions, there are two main sources of magnetic frustration: geometric frustration coming from the lattice structure itself, and exchange frustration originating from anisotropic spin interactions. By generating strong quantum fluctuations, both types of frustration can stabilize a quantum spin liquid (QSL) state[1], which offers a unique insight into how collective quantum behavior deviates from classical intuition[2]. While geometric frustration is commonly discussed in the context of non-bipartite (e.g., triangular, kagome, and pyrochlore) lattices, the most well-known example for exchange frustration is the Kitaev model on the honeycomb lattice, whose ground state is an exactly solvable QSL characterized by emergent gauge fluxes and Majorana fermions as elementary excitations[3].

The $S = 1/2$ triangular lattice antiferromagnet (TLAF) provides a quintessential manifestation of geometric frustration as the three antiferromagnetic interactions surrounding each triangle cannot be simultaneously satisfied. In the presence of both nearest-neighbor ($J_1$) and next-nearest-neighbor ($J_2$) Heisenberg interactions, this system is suggested to host a QSL ground state[4-10], which is notably stabilized by a very small $J_2 \approx 0.06 J_1$ (i.e., QSL arises for $0.06 < J_2/J_1 < 0.16$, see Fig. 1). Consequently, even though the purely nearest-neighbor Heisenberg TLAF ($J_2 = 0$) develops 120° magnetic order, its spin dynamics are still heavily influenced by quantum fluctuations originating from the proximate QSL phase. In particular, the measured excitation spectrum, which contains a strong continuum, cannot be reproduced by spin-wave theory even with non-linear ($1/S$) corrections[11-14]. Instead, it has been accurately modeled through a Schwinger boson approach based on fractionalized excitations (spinons) of the nearby QSL[15,16]. In this framework, the 120° ordered phase and its low-energy magnons are described through a spontaneous Bose-Einstein condensation of spinons and bound states of spinon pairs, respectively (i.e., spinons are confined). However, due to the proximity to a quantum melting point, a phase boundary of the QSL, spinons exhibit markedly less confinement at higher energies and freely propagate across multiple lattice spacings. Consequently, they manifest as an intense multi-spinon continuum above the one-magnon bands. Notably, this scenario is in stark contrast to unfrustrated two-dimensional quantum antiferromagnets, where the interacting magnon picture usually provides a quantitatively correct description of the excitation spectrum[17,18].

In the presence of anisotropic spin interactions, the $S = 1/2$ TLAF serves as a rich environment for combining geometric frustration with exchange frustration. Recent theoretical efforts have studied a general symmetry-allowed spin model for the $S = 1/2$ TLAF and revealed a rich phase diagram with intriguing phases not accessible by the purely isotropic (i.e., Heisenberg) model[19,20]. However, experimental exploration of this spin model is still in its early stages due to the lack of suitable $S = 1/2$ TLAF materials. Surprisingly, this challenge persists even for the planar XXZ anisotropy, which, among the various symmetry-allowed anisotropic exchange interactions, stands out as the simplest yet fundamentally important component:

$$H = \sum_{n,\langle ij \rangle_n} J_n \left( S_i^x S_j^x + S_i^y S_j^y + \Delta S_i^z S_j^z \right). \tag{1}$$

Here, $J_n$ represents the interactions between $n^{\text{th}}$ nearest neighbors, and $\Delta < 1$ denotes the degree of the XXZ anisotropy. This anisotropy smoothly transforms the Heisenberg model ($\Delta = 1$) into the XY model ($\Delta = 0$), each constituting its own universality class of phase transitions with distinct properties (e.g., Berezinskii–Kosterlitz–Thouless transition[21,22]). Thus, the planar XXZ anisotropy is an integral part of many foundational models in quantum magnetism and has been extensively studied in, for example, one-dimensional spin chains[23].

In a $S = 1/2$ TLAF, the planar XXZ anisotropy ($\Delta < 1$) is expected to suppress the inherent quantum fluctuations, according to recent theoretical works[19,20,24]. This is evident in the phase diagram of the $J_1$–$J_2$–$\Delta$ model (see Fig. 1) calculated by density matrix renormalization group (DMRG)[19,24]. As $\Delta$ deviates from 1, the QSL phase that separates the 120° and stripe magnetic orders is gradually suppressed and completely vanishes around $\Delta \sim 0.3$. This suggests that a magnon-based description of the spin dynamics may become more accurate in an $S = 1/2$ TLAF for small enough $\Delta$. However, as shown in Fig. 1, such an effect of the XXZ anisotropy on $S = 1/2$ TLAFs has yet to be investigated experimentally. Even though the XXZ model has already been applied to some $S = 1/2$ TLAF experimental systems such as $Ba_3CoSb_2O_9$ (Refs. [16,25]), the anisotropy is almost negligible in these materials ($\Delta \geq 0.9$) and thus inadequate to test the above hypothesis. In contrast, for $S = 1/2$ TLAFs with Ising-type XXZ anisotropy ($\Delta > 1$), recent experimental studies have unveiled fascinating static and dynamic magnetic properties associated with strong quantum fluctuations of the suggested spin supersolid phase[26-28]. This further calls for an experimental investigation into the case with strong planar XXZ anisotropy as it will enable us to understand the connection between the XY, Heisenberg, and Ising limits in $S = 1/2$ TLAFs, each exhibiting markedly different physical behaviors.

Here, we make a significant stride toward understanding the interplay between planar XXZ anisotropy and geometrical frustration in a $S = 1/2$ TLAF. We use inelastic neutron scattering (INS) and advanced theoretical calculations to study the magnetic excitations of $Ba_2La_2CoTe_2O_{12}$ (BLCTO), a $Co^{2+}$-based $S = 1/2$ TLAF that exhibits long-range 120° order on the $a$–$b$ plane below $T_N$ = 3.26 K [29]. First, by analyzing the excitation spectrum at $T > T_N$ with a state-of-the-art calculation technique, we establish that the spin Hamiltonian of BLCTO contains a substantial planar XXZ anisotropy: $\Delta$ = 0.56(2). Second, we present evidence of strong quantum fluctuations in BLCTO; the spectrum below $T_N$ exhibits continuum excitations up to approximately four times the single-magnon bandwidth, and these excitations possess a larger spectral weight than the single-magnon modes. A comparison with two advanced theoretical calculations suggests that a framework based on fractionalized spinons yields better quantitative agreement with the measured excitation spectrum than that of interacting magnons, highlighting the pivotal role played by quantum fluctuations even in the highly anisotropic regime of the phase diagram in Fig. 1. We also highlight another distinctive aspect of the planar XXZ anisotropy in $S = 1/2$ TLAFs: the pronounced flatness of the single-magnon and continuum excitations, which can be related to a higher-order van Hove singularity (VHS) in the magnon spectrum near $\Delta \sim 0.6$.

**Results and Discussion**

BLCTO possesses a trigonal crystal structure ($R\bar{3}$ space group) with $Co^{2+}$ sites forming layered triangular lattices[29] (Fig. 2a). The large separation between adjacent $Co^{2+}$ triangular layers (~ 9.16 Å) implies highly two-dimensional magnetic behavior. The $Co^{2+}$ site is surrounded by six $O^{2-}$ ligands configuring octahedral coordination (Fig. 2b). As found in many other $Co^{2+}$ systems with a similar environment[14,30-32], spin-orbit coupling (SOC) lifts the degeneracy of the $T_1$ manifold of the $Co^{2+}$ ions in BLCTO, resulting in a Kramers' doublet ground state (Fig. 2c) that gives rise to a pseudospin-1/2 entity ($J_{eff}$ = 1/2). The most direct means of verifying the $J_{eff}$ = 1/2 picture in this class of materials is the observation of an excitation from $J_{eff}$ = 1/2 to $J_{eff}$ = 3/2 multiplets (Fig. 2c)[30,31,33,34]. Our INS measurements at 5 K (above $T_N$) clearly show this transition (see Fig. 2d); an excitation corresponding to this transition is observed at ~28 meV, consistent with the energy scale of the single-ion $Co^{2+}$ SOC strength ($\lambda_{SOC}$ ~ 26 meV). Thus, BLCTO can be effectively treated as an $S = 1/2$ triangular lattice system when only the ground state doublet is thermally populated, i.e., for temperatures $T < 100$ K satisfying $3k_BT < E_{1/2 \to 3/2}$, where $E_{1/2 \to 3/2}$ is the transition energy from $J_{eff} = 1/2$ to $J_{eff} = 3/2$. Several hallmark characteristics of a $S = 1/2$ TLAF have been identified in BLCTO, including a substantial frustration factor $f = \theta_{CW}/T_N \sim 30$ (Fig. 2f–g), a field-induced 1/3–magnetization plateau phase[29], and a total magnetic entropy close to $R\ln2$ as estimated from heat capacity measurements (Fig. 2h–i).

However, as we show in this work, what sets BLCTO apart from other $S = 1/2$ TLAFs is its significant planar XXZ anisotropy. In general, a strong coupling between spin and orbital degrees of freedom in a $J_{eff} = 1/2$ ground state facilitates the emergence of strong anisotropy in exchange interactions[35,36]. In BLCTO, the $J_{eff} = 1/2$ ground state coexists with a trigonal distortion of the octahedral environment (Fig. 2b), which can generate sizable XXZ anisotropy [37,38]. The proportional relationship between XXZ anisotropy and trigonal distortion in a $Co^{2+}$ magnet, each quantified by $\Delta$ and the deviation of an O-Co-O angle from 90º (See Supplementary Fig. 1), is evident in Supplementary Table 1. BLCTO exhibits an exceptionally large trigonal compression of $CoO_6$ octahedra[29] compared to the other compounds (Fig. 2b). This is consistent with the fact that the measured $E_{1/2 \to 3/2}$ is much smaller than its theoretical value $3\lambda_{soc}/2 \sim 39$ meV without the distortion, implying the presence of a large trigonal crystal field[33,34] (Fig. 2e). For a more detailed explanation, see Supplementary Notes. These observations collectively make BLCTO a compelling case for strong XXZ anisotropy and, therefore, an attractive model system to study the unexplored region of the phase diagram shown in Fig. 1.

The strong quantum fluctuations present in $S = 1/2$ TLAFs disqualify many of the conventional methods for determining a material's spin Hamiltonian. Specifically, attempts to fit the magnetic excitations in the ordered phase ($T < T_N$) using linear spin-wave theory (LSWT) or non-linear spin-wave theory (NLSWT) do not produce reliable parameters. To overcome this challenge, we have examined the spin Hamiltonian of BLCTO by analyzing its excitation spectra measured above $T_N$. In this regime, the spin dynamics exhibit a semi-classical behavior owing to substantial thermal fluctuations. We analyzed the momentum- and energy-resolved paramagnetic scattering profile using Landau-Lifshitz dynamics (LLD) simulations (see Methods). This methodology provides more comprehensive insights than the traditional energy-integrated diffuse scattering analysis based on the self-consistent Gaussian approximation (SCGA).

The energy-resolved excitation spectra of paramagnetic BLCTO unambiguously reveal the presence of strong planar XXZ anisotropy. Fig. 3a shows momentum- and energy-resolved excitation spectra of polycrystalline BLCTO measured at four temperatures above $T_N$. We first compare the entire dataset to LLD simulations with ($\Delta = 0.56$, Fig. 3b) and without ($\Delta = 1$, Fig. 3c) XXZ anisotropy. These two models yield distinct energy-dependent profiles, with the main differences highlighted in Fig. 3d which shows the overall energy dependence for $T = 1.63T_N$ by integrating over multiple magnetic zones in the powder spectrum. The isotropic model ($\Delta = 1$) fails to replicate the concentration of the INS cross-section in the low-energy region, and this discrepancy is remedied by reducing $\Delta$ (see Supplementary Fig. 4). For more quantitative analysis, we conducted a brute-force comparison between the entire dataset in Fig. 3a and the INS cross-sections ($I(|\mathbf{q}|,\omega)$) calculated from LLD across a wide two-dimensional parameter space: 2 meV < $J_1$ < 2.9 meV and 0.3 < $\Delta$ < 1.0, assuming $J_2 = 0$ (see Methods). The result, depicted in Fig. 3e, reveals a clear minimum of the chi-square between the observed and simulated INS cross-sections ($\chi^2(J_1, \Delta)$) at $J_1 = 2.62$ meV and $\Delta = 0.56(2)$. Indeed, the corresponding LLD simulation result (Fig. 3b) shows an excellent agreement with the data, whose detailed comparison can be found in Supplementary Fig. 5. This optimal parameter set clearly indicates the presence of strong planar XXZ anisotropy in BLCTO, in stark contrast to other previously reported examples of $S = 1/2$ TLAFs (see Fig. 1). The same conclusion is derived from our additional analysis based on the energy-integrated diffuse scattering profile, as shown in Supplementary Fig. 9.

Importantly, this result remains robust even when the analysis is extended to include next-nearest-neighbor interactions $J_2$ (see Supplementary Note and Supplementary Figs. 6–8) and bond-dependent anisotropic exchange interactions (see Supplementary Note and Supplementary Fig. 15). Notably, $J_2$ has a stronger impact on the $|\mathbf{q}|$-dependence of $I(|\mathbf{q}|,\omega)$, while the effect of $\Delta$ is more pronounced along the $\omega$–axis (Supplementary Fig. 4–6). The optimal parameter set obtained from the $J_1$–$J_2$–$\Delta$ model ($J_1 = 2.70(5)$ meV, $\Delta = 0.54(3)$, $J_2 = 0.03(1)J_1$) is marked as a blue star in Fig. 1. Another noteworthy component is the inter-layer coupling ($J_c$), which was suggested to be ferromagnetic based on the interlayer spin configuration confirmed by neutron diffraction[29]. Yet the layered crystal structure (Fig. 2a) and a broad maximum of the temperature-dependent magnetization (Fig. 2f) suggest strong 2D magnetic character in BLCTO, i.e., $J_c$ should be much smaller than $J_1$. In turn, adding such a small $J_c$ will only result in a marginal change of a powder-averaged excitation spectrum (see Supplementary Fig. 16).

Having established the presence of strong planar XXZ anisotropy in BLCTO, we now examine the spin dynamics in the ordered phase to determine the impact of quantum fluctuations. Notably, this provides, to our knowledge for the first time, experimental insights into the nature of magnetism in a 120°-ordered $S = 1/2$ TLAF with strong planar XXZ anisotropy. Figs. 4a and 4e show the measured excitation spectra of BLCTO at 1.8 K ($< T_\text{N}$ = 3.26 K). In addition to the one-magnon spectrum located below ~2 meV, the data reveal remarkable features at higher energies. A substantial region of the |**Q**|– $E$ space above the one-magnon region is covered by a continuum signal, extending up to roughly four times the one-magnon bandwidth; see Fig. 4i. Notably, the energy-dependent profile of Fig. 4e demonstrates a multi-level structure inside the continuum with two local maxima centered at around 2.2 and 3.1 meV (Fig. 4j), respectively designated by purple and red arrows throughout Fig 4. Most importantly, the spectral weight of this high-energy continuum surpasses that of the one-magnon band: $\frac{I_\text{highE}}{I_\text{1-mag}}$ ~1.34, according to Fig. 4i. Given that the fine quality of our sample (see Methods) reasonably excludes structural disorder as a contributing factor on the continuum, this observation suggests that quantum fluctuations play a significant role in the $S = 1/2$ TLAF with strong planar XXZ anisotropy.

For a deeper understanding, we compare our data with NLSWT and Schwinger boson calculations, which are theoretical approaches based on interacting magnons and fractionalized spinons, respectively. Our Schwinger boson approach employs a 1/$N$ expansion centered around a saddle point (SP) solution, where $N$ represents the number of boson flavors. Although the SP solution captures the 120° magnetic order through spontaneous Bose-Einstein condensation of Schwinger bosons, it cannot account for collective modes of this magnetic order, which are magnons. This limitation arises because the SP Hamiltonian models a non-interacting spinon gas, whereas magnons emerge as bound states of two spinons. Thus, by incorporating the 1/$N$ correction, which acts as a counter-diagram for the single-spinon poles of the SP dynamical spin susceptibility and captures the single-magnon poles as two-spinon bound states[15,16,39] (see Supplementary Fig. 18), we achieve a spectrum consisting of both low-energy single-magnon modes and a two-spinon scattering continuum above them. Upon incorporating a higher order of the 1/$N$ expansion (1/$N^2$), an effective two-magnon continuum will also appear as the pair creation of two-spinon bound states, extending to twice the highest energy of the two-spinon bound states. However, the spectral weight of this two-magnon continuum is expected to be smaller than that of the multi-spinon continuum.

While the Schwinger boson theory is known to better describe the excitation spectra of $S = 1/2$ Heisenberg TLAFs (Ref. [15,16,39]), one may expect NLSWT to be a more suitable approach in the XY limit, as described in the Introduction. Thus, it is not a priori clear which approach better agrees with the excitation spectrum of BLCTO, and comparing the results of each calculation with the BLCTO data reveals how important quantum fluctuations are in the spin dynamics of $S = 1/2$ TLAFs with strong planar XXZ anisotropy. For each calculation, we used $\Delta = 0.56(2)$ determined from our analysis of the high-temperature spin dynamics when $J_2 = 0$ (Fig. 3). However, adjusting $J_1$ to be between 1.55 and 1.77 meV (depending on the calculation method, see Figs. 4b–d and Supplementary Fig. 13) was necessary to align with the overall scale of our INS data measured below $T_N$. This modification is due to the inability of our semi-classical LLD simulation to account for the effective magnetic moment $\mu_{\text{eff}} \sim g(S(S+1))^{1/2}$ at high temperature [40]. Consequently, the LLD result overestimates $J_1$ by a factor of at most $\left(\frac{S(S+1)}{S^2}\right)^{1/2}$ (i.e., about 1.732 for $S = 1/2$), which is comparable to the ratio between the values of $J_1$ obtained from our high- and low-temperature analyses. Indeed, the smaller $J_1$ values derived at $T < T_N$ better align with the experimental $M$–$H$ curve of BLCTO [29] (see Supplementary Notes). We also note that anisotropic $g$-factors are taken into account in the calculations for both $T > T_N$ and $T < T_N$, which can substantially change the distribution of theoretical INS cross-sections (see Methods and Supplementary Notes).

Figs. 4c–d (4g–h) show the NLSWT and Schwinger boson calculation results convoluted with the instrumental resolution profile for $E_i = 10$ meV (6 meV). While both calculations qualitatively capture the multi-level structure at 1.5, 2.2, and 3.1 meV (Fig. 4i–l), detailed features in the measured spectrum (e.g., an intense flat mode around 1.5 meV, see Fig. 4e) are better described by the Schwinger boson calculation, as seen in Fig. 4h. Most importantly, these two calculations exhibit distinct spectral-weight ratios of the high-energy continuum to the one-magnon component: $\frac{I_{\text{highE}}}{I_{1-\text{mag}}}$. Based on the plots shown in Figs. 4i–j, we obtained a spectral-weight ratio between 0.22 and 0.25 for the NLSWT calculation and between 1.49 and 1.68 for the Schwinger boson calculation. Indeed, NLSWT significantly underestimates the spectral weight of the continuum above the one-magnon band, as evident in Figs. 4i–j and 4m–n. In contrast, the Schwinger boson theory captures $\frac{I_{\text{highE}}}{I_{1-\text{mag}}} > 1$ and the intensity modulation along $|\mathbf{Q}|$ (Fig. 4m–o) observed in the data. Note that the spectral weight above ~4 meV found in the data (approximately 21% of the total spectral weight), which is not captured by the Schwinger boson theory, is due to the presence of higher-order $n$-spinon states with $n > 2$, which are not implemented in our current calculation[16]. Including these terms in the calculation is expected to transfer the overestimated spectral weight around the 3rd level (red arrow in Fig. 4j) to higher energies and better describe the INS data. To summarize, the spinon-based picture underpinning

Schwinger boson theory is found to be a better framework for reproducing the excitation spectrum of BLCTO, suggesting that spinons are fundamental ingredients for describing the spin dynamics of $S = 1/2$ TLAFs with significant planar XXZ anisotropy. This is presumably because a quantum melting point remains proximate to the 120° ordered phase even under strong planar XXZ anisotropy, which may largely weaken the confinement of spinons. Thus, the noticeable suppression of quantum fluctuations by planar XXZ anisotropy is expected to be found at $\Delta$ much smaller than ~0.5. Nevertheless, it is noteworthy that the Schwinger boson calculation overestimates the $\frac{I_{highE}}{I_{1-mag}}$ ratio found in our data, which signals that BLCTO is still less quantum than Heisenberg TLAFs, as anticipated from the phase diagram in Fig. 1.

We now turn our attention to another interesting feature of the magnetic excitations that results from strong planar XXZ anisotropy. In addition to the intense continuum signal, the 1st level (grey arrows in Fig. 4) and 2nd level (purple arrows in Fig. 4) excitations exhibit a flat profile along the momentum transfer axis. In the language of magnons, the flat continuum signal can be interpreted as originating from the flatness of the single-magnon dispersion within a portion of the Brillouin zone. Since there are many magnons at the approximately constant energy $E$ of this flat region, there are many channels for creating two magnons with total energy $2E$, resulting in a pronounced peak in the two-magnon density of states at energy $2E$. This rationale is evidenced in Supplementary Fig. 11: when the overall one-magnon dispersion becomes noticeably flat ($\Delta < 0.5$), the NLSWT spectrum manifests a flat, isolated profile of a two-magnon continuum.

Understanding the origin of the wide flat region in the single-magnon dispersion reveals a unique role played by the planar XXZ anisotropy. To see this, it is useful to describe the magnons in a local reference frame where the spins are rotated in the plane of the 120° order such that they all point in the same direction[41,42]. In this rotating reference frame, the magnon spectrum has only one band, which supports a Goldstone mode at the Γ point for any value of the planar XXZ anisotropy $\Delta$. While the K points also host Goldstone modes in the Heisenberg limit ($\Delta = 1$), they become gapped for a generic value of $\Delta$ and even evolve into local band maxima in the XY limit ($\Delta = 0$). Hence, as $\Delta$ is progressively reduced, each K point must undergo a transition from a local minimum to a local maximum at a critical value $\Delta = \Delta_0$. Due to the high symmetry of each K point, this transition at the critical anisotropy $\Delta_0$ must happen through a higher-order VHS, at which all first and second derivatives of the magnon band dispersion vanish and the magnon density of states exhibits a power-law divergence $\rho(\varepsilon) \sim |\varepsilon - E|^{-1/3}$ [43-46]. To show this, we consider the general symmetry-allowed Taylor expansion of the magnon dispersion up to the third order around a given K point. If we introduce the relative momentum $(k_x, k_y)$, the threefold rotation symmetry around (0, 0) (i.e., the K point itself) and the reflection symmetry $k_y \rightarrow -k_y$ imply that this Taylor expansion takes the form:

$$E(k_x, k_y) = E(0, 0) + c_2 (k_x^2 + k_y^2) + c_3 (k_x^3 - 3 k_x k_y^2). \qquad (2)$$

Clearly, the transition between local minimum and local maximum is driven by a sign change of the lowest order nonconstant term, i.e., the second-order term of coefficient $c_2$. Since $c_2 = 0$ at the transition ($\Delta = \Delta_0$), the lowest order magnon dispersion around the K point is then determined by the third order term, which precisely corresponds to the higher order VHS (with a shape referred to as a *monkey saddle*) discussed in, for example, Ref. [43]. The magnon dispersions on the two sides of the transition and at the transition itself are shown in Fig. 5. We also emphasize that, by returning from the rotating reference frame to the laboratory frame, two additional shifted copies of the original magnon dispersion arise in energy-momentum space, resulting in flat regions coming from the higher-order VHS not only around the K points but also around the Γ point. The above arguments derived from Eq. (2) are independent of the models for elementary excitations, i.e., they hold for LSWT, NLSWT, and Schwinger Boson calculations alike. Indeed, all three different calculation methods consistently suggest that planar XXZ anisotropy enhances the flatness of one-magnon bands, and the flatness is maximized at appropriate values of $\Delta$ much lower than 1, which corresponds to $\Delta_0$ (Supplementary Figs. 10–12 and 19).

Interestingly, the experimentally determined anisotropy $\Delta$ of BLCTO (0.56(2) for the $J_1$–$\Delta$ model) is not far from the theoretical critical anisotropy $\Delta_0$. While LSWT and NLSWT give $\Delta_0 \approx 0.25$ and 0.35, respectively, significantly smaller than the experimental value for BLCTO, Schwinger boson theory, which has been found to be the most suitable approach for computing the spin dynamics, gives $\Delta_0 \approx 0.6$ (see Supplementary Fig. 14). Indeed, our Schwinger boson calculations corresponding to the case of BLCTO ($\Delta = 0.56$) show that the nearly flat dispersion at ~1.5 meV persists along the entire high-symmetry line Γ–K–M (see the single-crystal spectrum in Fig. 4l). However, the dispersion around the Γ and K points becomes less flat when moving away from $\Delta_0 \approx 0.6$ (Supplementary Fig. 14), resulting in a less flat powder-averaged spectrum (Supplementary Fig. 12). Thus, these flat excitations originating from a higher-order VHS are unique features of the strongly anisotropic XXZ model realized in BLCTO, and provide a clear distinction from both the XY ($\Delta = 0$) and Heisenberg ($\Delta = 1$) limits.

Finally, we discuss the validity and limitations of our spin Hamiltonian. Although we proposed a relatively simple $J_1$–$J_2$–$\Delta$ model for the spin Hamiltonian of BLCTO (or approximately, the $J_1$–$\Delta$ model as $J_2$ is very small), bond-dependent anisotropic interactions ($J_{\pm\pm}$ and $J_{z\pm}$ in Ref. [20]), which transform into Kitaev ($K$) and Gamma ($\Gamma$) terms in local cubic axes[19,20], are in principle allowed in BLCTO by symmetry. Notably, $Co^{2+}$ systems have proven to be an excellent platform for discovering significant bond-dependent anisotropic exchange interactions[33,34,47,48]. Although the isolated $CoO_6$ octahedra in BLCTO differ from previously reported Kitaev materials that consist of an edge-shared octahedron network of ligands (this is an important prerequisite of the Jackeli-Khaliullin mechanism[49]), a possibility of Kitaev interactions in an isolated octahedron network was suggested theoretically[50,51]. Thus, we additionally analyzed the paramagnetic excitation profiles with the $J_1$–$J_2$–$\Delta$–$J_{\pm\pm}$–$J_{z\pm}$ model. While single-crystal INS measurements will be necessary to quantify $J_{\pm\pm}$ and $J_{z\pm}$ with better accuracy, our analysis suggests that $J_{\pm\pm}$ and $J_{z\pm}$ are not significant (see Supplementary Notes and Supplementary Fig. 15). The correlation between this outcome and the existence of a nearly gapless Goldstone mode (Supplementary Fig. 17) also merits more discussion, which is included in Supplementary Notes.

In summary, we have investigated the spin dynamics of BLCTO using INS, marking to our best knowledge the first comprehensive experimental study of a $S = 1/2$ TLAF with strong planar-type XXZ anisotropy. We implemented a new reliable method to extract the spin Hamiltonian of a $S = 1/2$ TLAF from energy- and momentum-resolved spin dynamics at $T > T_N$, which suggests an XXZ anisotropy of $\Delta \sim 0.56(2)$ in BLCTO. The full excitation spectrum below $T_N$ reveals an intense and structured continuum signal distributed up to an energy around four times higher than the top of the single-magnon band. This observation, coupled with comparisons to NLSWT and Schwinger boson calculations, supports the persistence of strong quantum fluctuations and indicates that the anomalous continuum scattering is more accurately described by a two-spinon continuum even in the highly anisotropic regime of a $S = 1/2$ TLAF. The observed flatness of both single-magnon and continuum excitations, along with the associated higher-order van Hove singularity predicted by our calculations, further highlights strong planar XXZ anisotropy as an intriguing element in the generalized spin model of $S$ = 1/2 TLAFs. Our results offer a rare experimental perspective on a highly anisotropic $S$ = 1/2 TLAF that bridges the Heisenberg and XY models.

## Methods

**Sample preparation.**

Poly-crystalline Ba$_2$La$_2$CoTe$_2$O$_{12}$ (BLCTO) samples were prepared by the solid-state reaction of BaCO, (99.98%, Sigma Aldrich), La$_2$O$_3$ (99.99%, Sigma Aldrich), CoO (99.995%, Alfa Aesar), and TeO$_2$ (99.995%, Sigma Aldrich). They were mixed in stoichiometric quantities by grinding in an agate mortar to produce a starting charge of approximately 15 grams of powder. The charge was placed in a 50 ml platinum crucible with a lid and sintered under ambient air condition at 1100°C for 36 hours using an electric resistance box furnace. The heating rate was 100°C per hour. After heating for 36 hours, the charge was cooled down to room temperature and the resulting medium purple powder was reground in an agate mortar and pressed in a 20 mm diameter stainless steel die to produce two pellets. The pellets were placed into the covered platinum crucible and sintered again under identical conditions as the first sintering sequence. The resultant material with a uniform purple color was characterized by powder X-ray diffraction, which confirmed the formation of BLCTO. Further characterization using powder neutron diffraction (see the relevant subsection below) verified high purity and quality of the sample, as displayed in Supplementary Fig. 2 and Supplementary Table 2.

**Bulk property measurements.**

The magnetic susceptibility of BLCTO was measured by a commercial magnetometer MPMS-XL (Quantum Design, USA). Following the previous reference[52], we used a modified Curie-Weiss law to fit our $M$(T) data:

$$\chi(T) = \chi_0 + \frac{C}{T - \theta_{\text{CW}}}, \tag{3}$$

where $\chi_0$ is an additional van Vleck-like term attributed to the non-negligible population of the $J_{\text{eff}} = 3/2$ state (see Fig. 2d) at high temperatures. The fitting results for temperatures between 100 K and 150 K are $\chi_0 = 4.5(45) \times 10^{-4}$ emu mol$^{-1}$ Oe$^{-1}$, $C = 3.73(21)$ emu mol$^{-1}$ Oe$^{-1}$ K$^{-1}$, and $\theta_{\text{CW}} = -99(6)$ K; see orange solid lines in Fig. 2f–g.

Heat capacity was measured by using a Quantum Design instrument. The sample consisted of a 4.75 mg pressed pellet held to the calorimeter stage with grease. A non-magnetic (= phonon) contribution to our measured data was estimated by fitting the range of 40 ~ 200 K to the simplified Debye model. We used two different Debye temperatures for the fitting, which is justified by the co-existence of heavy (La, Te) and light (O) elements. This comparison yields $\theta_{\text{D1}} = 237(2)$ K and $\theta_{\text{D2}} = 746(3)$ K. While this resulted in a reasonable estimate (Fig. 2i), we acknowledge the potential inaccuracies due to the more complex nature of the actual phonon density-of-states spectrum in BLCTO. Both the magnetic susceptibility and heat capacity results are consistent with Ref. [29].

**Powder neutron diffraction**

For crystal structure characterization, we conducted neutron powder diffraction measurements on the high-resolution powder diffractometer (HB-2A) at the High Flux Isotope Reactor (HFIR), ORNL. The data were collected at 200 K using a constant incident neutron wavelength of 1.539 Å. Rietveld refinements were performed using the FullProf software package [53].

**Powder inelastic neutron scattering (INS).**

We performed two INS measurements using 10.4 grams of powder BLCTO sealed with He exchange gas within a thin-walled cylindrical aluminum sample container. The spin-orbit exciton signal was obtained from the SEQUOIA time-of-flight spectrometer at the Spallation Neutron Source (SNS), ORNL. The data were collected at 5, 20, 100, 200, and 300 K with incident neutron energies of $E_i$ = 40 meV and 65 meV with standard high-resolution chopper conditions. The low-energy magnetic excitation data were collected at the CNCS time-of-flight spectrometer at the SNS, ORNL. The data were collected at 1.65 K, 5.3 K, 5.6 K, 20.2 K, and 40.4 K using multiple incident neutron energies ($E_i$ = 1.5, 3.3, 6.0, and 10.0 meV for 1.65 K, $E_i$ = 6.0 meV for 5.3 K, and $E_i$ = 10 meV for 5.6, 20.2, and 40.4 K) with standard high-flux chopper conditions. A background signal for each measurement was acquired by measuring a separate empty holder. Unless otherwise described, these data were subtracted from our main data. The instrumental energy resolution was modeled by utilizing the *Pychop* package in *Mantid*[54]. The momentum resolution of each dataset was estimated from the full width at half-maximum (FWHM) of the Bragg peaks of BLCTO and/or aluminum. As a result, we obtained dQ = 0.036 Å$^{-1}$, 0.044 Å$^{-1}$, and 0.06 Å$^{-1}$ for $E_i$ = 3.3, 6.0, and 10.0 meV, respectively. The data cuts along energy transfer $E$ (e.g. Fig. 3d and Fig. 4i–j) were obtained by averaging dynamical structure factor values at the same $E$ but different |**Q**|. We used the same averaging process for the data cuts along |**Q**|.

**Calculating and fitting paramagnetic excitations.**

To calculate the energy-resolved magnetic excitations of paramagnetic BLCTO, we performed a Landau-Lifshitz dynamics (LLD) simulation with the spin system governed by the spin Hamiltonian in Eq. (1). This was done by using the LLD simulation package Su(n)ny[55], whose detailed working principles can be found in Refs. [56-58]. We used a 45×45×2 supercell of BLCTO (12,150 Co$^{2+}$ sites) for the simulation.

To implement the same degree of thermal fluctuations as the measured data (Fig. 3a) in LLD simulations, we set the simulation temperature of LLD through the following steps. First, we normalized the experimental temperatures by the material's $T_{N, real}$ = 3.26 K (e.g., 5.3 K = 1.63 $T_{N, real}$). Then, we computed the theoretical $T_N$ of a spin system as a function of exchange parameters (e.g. $T_N(J_1, J_2, \Delta)$ for the $J_1$–$J_2$–$\Delta$ model) across the entire range of the three-dimensional ($J_1$, $J_2$, $\Delta$) parameter space to be investigated. This calculation was done using a separate classical Monte-Carlo simulation based on LLD and simulated annealing. In this simulation, a Langevin time step d$t$ and the damping constant λ were set to 0.11 meV$^{-1}$ and 0.1, respectively. After waiting the 1,000 Langevin time steps for equilibration, the 80,000 Langevin time steps were used for the sampling. The effective $T_N$ of the simulation was determined from the temperature dependence of the calculated heat capacity, which in turn was derived from the variance of the total energy values sampled throughout the 80,000 time steps. Using the obtained $T_N(J_1, J_2, \Delta)$ and the experimental temperature normalized by $T_{N, real}$, we assigned parameter-dependent simulation temperatures. For instance, we used $T = 1.63T_N(J_1, J_2, \Delta)$ for the simulation corresponding to the data measured at 5.3 K = 1.63$T_{N,real}$.

For the LLD simulation to calculate dynamical structure factors, a Langevin time step was set to d$t$ = 0.11 meV$^-$

$^1$, 0.11 meV$^{-1}$, 0.04 meV$^{-1}$, and 0.07 meV$^{-1}$ for simulations at $T$ = 1.63$T_N$, 1.72$T_N$, 6.20 $T_N$, and 12.40 $T_N$, respectively. The damping constant $\lambda$ was fixed at 0.1. The first 1,500 Langevin time steps were discarded to wait for equilibration. The calculated dynamical structure factor $S(\mathbf{Q},\omega)$ was converted to INS cross-sections $I(|\mathbf{Q}|,\omega)$ through a process involving powder averaging, applying the magnetic form factor of Co$^{2+}$, applying neutron polarization factors ($1 - Q_iQ_j/|\mathbf{Q}|^2$ for $S_{ij}(\mathbf{Q}, \omega)$), with an anisotropic $g$-tensor (see Supplementary Notes and Supplementary Fig. 3) and instrumental resolution convolution. Note that the same treatments were applied to our linear/non-linear spin-wave calculations and Schwinger Boson calculations. To ensure statistically reliable outcomes, we repeated the simulation 10 times and used the averaged $I(|\mathbf{Q}|,\omega)$ for comparisons with the measured $I(|\mathbf{Q}|,\omega)$.

The goodness-of-fit between the data and LLD simulations was quantified by the chi-square ($\chi^2$):

$$\chi^2 = \sum_{i,j} \frac{\left[I_{\exp}(|\mathbf{Q}|_i,\omega_j) - \left(b + fI_{\mathrm{calc}}(|\mathbf{Q}|_i,\omega_j)\right)\right]^2}{(\sigma_{ij})^2}, \tag{4}$$

where $i$ and $j$ are indices for data points in the $|\mathbf{Q}|$–$\omega$ space, $I_{\exp}(|\mathbf{Q}|, \omega)$ and $I_{\mathrm{calc}}(|\mathbf{Q}|, \omega)$ are the measured and calculated INS cross-sections, and $\sigma_{ij}$ is the standard deviation of $I_{\exp}(|\mathbf{Q}|_i, \omega_j)$. $f$ and $b$ are a scale and constant background parameter that are refined to minimize $\chi^2$ for given $I_{\exp}(|\mathbf{Q}|, \omega)$ and $I_{\mathrm{calc}}(|\mathbf{Q}|, \omega)$ maps, which can be uniquely determined by using Eq. 4 in Ref. [59]. For this $\chi^2$ analysis, we considered the four INS data shown in Fig. 3a simultaneously with different values of $f$ and $b$ for each effective temperature. The masked regions in Fig. 3a (white areas), except for those outside the energy-momentum coverage of time-of-flight powder neutron spectroscopy, were excluded from the analysis as they contain spurious signals or a quasi-elastic background signal.

**Spin-wave and Schwinger Boson calculations.**
Linear spin-wave theory calculations were performed using the SpinW library[60]. Non-linear spin-wave theory calculations were made based on 1/$S$ corrections. The first 1/$S$ corrections to linear spin-wave theory in both the magnon energies and the dynamical spin structure factor are obtained by the standard approach described in Refs. [41] and [42]. The only differences are that (a) the magnon energies are evaluated in the on-shell approximation and (b) the first 1/$S$ corrections are included in both the one-magnon and the two-magnon energies (see also Ref. [61]). The explanation of Schwinger Boson calculations is provided in Supplementary Notes.

## Data Availability

The source data used in Fig. 1 are available in the figshare database under accession code https://doi.org/10.6084/m9.figshare.26490433. The authors declare that other data supporting the findings of this study are available within the paper and the Supplementary Information. Further raw data are available from the corresponding author upon request.

## Code Availability

Custom codes used in this article are available from the corresponding authors upon reasonable request.

# References

1. Anderson, P. W. Resonating valence bonds: A new kind of insulator? *Materials Research Bulletin* **8**, 153-160 (1973).

2. Broholm, C. *et al.* Quantum spin liquids. *Science* **367**, eaay0668 (2020).

3. Kitaev, A. Anyons in an exactly solved model and beyond. *Annals of Physics* **321**, 2-111 (2006).

4. Zhu, Z. & White, S. R. Spin liquid phase of the S= 1/2 J1−J2 Heisenberg model on the triangular lattice. *Physical Review B* **92**, 041105 (2015).

5. Hu, W.-J., Gong, S.-S., Zhu, W. & Sheng, D. Competing spin-liquid states in the spin-1/2 Heisenberg model on the triangular lattice. *Physical Review B* **92**, 140403 (2015).

6. Iqbal, Y., Hu, W.-J., Thomale, R., Poilblanc, D. & Becca, F. Spin liquid nature in the Heisenberg J1−J2 triangular antiferromagnet. *Physical Review B* **93**, 144411 (2016).

7. Saadatmand, S. & McCulloch, I. Symmetry fractionalization in the topological phase of the spin-1/2 J1−J2 triangular Heisenberg model. *Physical Review B* **94**, 121111 (2016).

8. Wietek, A. & Läuchli, A. M. Chiral spin liquid and quantum criticality in extended s= 1/2 heisenberg models on the triangular lattice. *Physical Review B* **95**, 035141 (2017).

9. Gong, S.-S., Zhu, W., Zhu, J.-X., Sheng, D. N. & Yang, K. Global phase diagram and quantum spin liquids in a spin-1 2 triangular antiferromagnet. *Physical Review B* **96**, 075116 (2017).

10. Hu, S., Zhu, W., Eggert, S. & He, Y.-C. Dirac spin liquid on the spin-1/2 triangular Heisenberg antiferromagnet. *Physical review letters* **123**, 207203 (2019).

11. Verresen, R., Moessner, R. & Pollmann, F. Avoided quasiparticle decay from strong quantum interactions. *Nature Physics* **15**, 750-753 (2019).

12. Macdougal, D. *et al.* Avoided quasiparticle decay and enhanced excitation continuum in the spin-1 2 near-Heisenberg triangular antiferromagnet Ba3CoSb2O9. *Physical Review B* **102**, 064421 (2020).

13. Ma, J. *et al.* Static and dynamical properties of the spin-1/2 equilateral triangular-lattice antiferromagnet Ba3CoSb2O9. *Physical review letters* **116**, 087201 (2016).

14. Ito, S. *et al.* Structure of the magnetic excitations in the spin-1/2 triangular-lattice Heisenberg antiferromagnet Ba3CoSb2O9. *Nature communications* **8**, 235 (2017).

15. Scheie, A. O. *et al.* Proximate spin liquid and fractionalization in the triangular antiferromagnet KYbSe2. *Nature Physics* **20**, 74-81 (2024).

16. Ghioldi, E. *et al.* Evidence of two-spinon bound states in the magnetic spectrum of Ba3CoSb2O9. *Physical Review B* **106**, 064418 (2022).

17. Sala, G. *et al.* Van Hove singularity in the magnon spectrum of the antiferromagnetic quantum honeycomb lattice. *Nature communications* **12**, 171 (2021).

18. Coldea, R. *et al.* Spin Waves and Electronic Interactions in $La_2CuO_4$. *Physical Review Letters* **86**, 5377-5380 (2001).

19  Zhu, Z., Maksimov, P., White, S. R. & Chernyshev, A. Topography of spin liquids on a triangular lattice. *Physical Review Letters* **120**, 207203 (2018).

20  Maksimov, P. A., Zhu, Z., White, S. R. & Chernyshev, A. L. Anisotropic-Exchange Magnets on a Triangular Lattice: Spin Waves, Accidental Degeneracies, and Dual Spin Liquids. *Physical Review X* **9**, 021017 (2019).

21  Berezinskii, V. Destruction of long-range order in one-dimensional and two-dimensional systems having a continuous symmetry group I. Classical systems. *Sov. Phys. JETP* **32**, 493-500 (1971).

22  Kosterlitz, J. & Thouless, D. Ordering, metastability and phase transitions in two-dimensional systems 1973. *J. Phys.: Condens. Matter* **6**, 1181.

23  Šamaj, L. & Bajnok, Z. *Introduction to the statistical physics of integrable many-body systems*. (Cambridge University Press, 2013).

24  Zhu, Z., Maksimov, P., White, S. R. & Chernyshev, A. Disorder-induced mimicry of a spin liquid in YbMgGaO4. *Physical review letters* **119**, 157201 (2017).

25  Kamiya, Y. *et al.* The nature of spin excitations in the one-third magnetization plateau phase of Ba3CoSb2O9. *Nature communications* **9**, 2666 (2018).

26  Zhong, R., Guo, S., Xu, G., Xu, Z. & Cava, R. J. Strong quantum fluctuations in a quantum spin liquid candidate with a Co-based triangular lattice. *Proceedings of the National Academy of Sciences* **116**, 14505-14510 (2019).

27  Xiang, J. *et al.* Giant magnetocaloric effect in spin supersolid candidate Na2BaCo(PO4)2. *Nature* **625**, 270-275 (2024).

28  Zhu, M. *et al.* Continuum excitations in a spin-supersolid on a triangular lattice. *arXiv preprint arXiv:2401.16581* (2024).

29  Kojima, Y. *et al.* Quantum magnetic properties of the spin-1 2 triangular-lattice antiferromagnet Ba2La2CoTe2O12. *Physical Review B* **98**, 174406 (2018).

30  Kim, C. *et al.* Bond-dependent anisotropy and magnon decay in cobalt-based Kitaev triangular antiferromagnet. *Nature Physics* **19**, 1624-1629 (2023).

31  Yuan, B. *et al.* Spin-orbit exciton in a honeycomb lattice magnet CoTiO3: revealing a link between magnetism in d-and f-electron systems. *Physical Review B* **102**, 134404 (2020).

32  Kojima, Y., Kurita, N., Tanaka, H. & Nakajima, K. Magnons and spinons in Ba2CoTeO6: A composite system of isolated spin-1/2 triangular Heisenberg-like and frustrated honeycomb Ising-like antiferromagnets. *Physical Review B* **105**, L020408 (2022).

33  Liu, H. & Khaliullin, G. Pseudospin exchange interactions in d 7 cobalt compounds: Possible realization of the Kitaev model. *Physical Review B* **97**, 014407 (2018).

34  Liu, H., Chaloupka, J. & Khaliullin, G. Kitaev Spin Liquid in 3d Transition Metal Compounds. *Physical Review Letters* **125**, 047201 (2020).

35  Witczak-Krempa, W., Chen, G., Kim, Y. B. & Balents, L. Correlated Quantum Phenomena in the Strong Spin-Orbit Regime. *Annual Review of Condensed Matter Physics* **5**, 57-82 (2014).

36  Rau, J. G., Lee, E. K.-H. & Kee, H.-Y. Spin-Orbit Physics Giving Rise to Novel Phases in Correlated Systems: Iridates and Related Materials. *Annual Review of Condensed Matter Physics* **7**, 195-221 (2016).

37  Liu, X. & Kee, H.-Y. Non-Kitaev versus Kitaev honeycomb cobaltates. *Physical Review B* **107**, 054420 (2023).

38  Yuan, B. *et al.* Dirac Magnons in a Honeycomb Lattice Quantum XY Magnet CoTiO3. *Physical Review X* **10**, 011062 (2020).

39  Ghioldi, E. A. *et al.* Dynamical structure factor of the triangular antiferromagnet: Schwinger boson theory beyond mean field. *Physical Review B* **98**, 184403 (2018).

40  Dahlbom, D. *et al.* Quantum-to-classical crossover in generalized spin systems: Temperature-dependent spin dynamics of FeI2. *Physical Review B* **109**, 014427 (2024).

41  Chernyshev, A. & Zhitomirsky, M. Spin waves in a triangular lattice antiferromagnet: decays, spectrum renormalization, and singularities. *Physical Review B* **79**, 144416 (2009).

42  Mourigal, M., Fuhrman, W., Chernyshev, A. & Zhitomirsky, M. Dynamical structure factor of the triangular-lattice antiferromagnet. *Physical Review B* **88**, 094407 (2013).

43  Shtyk, A., Goldstein, G. & Chamon, C. Electrons at the monkey saddle: A multicritical Lifshitz point. *Physical Review B* **95**, 035137 (2017).

44  Yuan, N. F. Q., Isobe, H. & Fu, L. Magic of high-order van Hove singularity. *Nature Communications* **10**, 5769 (2019).

45  Chandrasekaran, A., Shtyk, A., Betouras, J. J. & Chamon, C. Catastrophe theory classification of Fermi surface topological transitions in two dimensions. *Physical Review Research* **2**, 013355 (2020).

46  Yuan, N. F. Q. & Fu, L. Classification of critical points in energy bands based on topology, scaling, and symmetry. *Physical Review B* **101**, 125120 (2020).

47  Kim, C., Kim, H.-S. & Park, J.-G. Spin-orbital entangled state and realization of Kitaev physics in 3d cobalt compounds: a progress report. *Journal of Physics: Condensed Matter* **34**, 023001 (2022).

48  Sano, R., Kato, Y. & Motome, Y. Kitaev-Heisenberg Hamiltonian for high-spin d7 Mott insulators. *Physical Review B* **97**, 014408 (2018).

49  Chaloupka, J., Jackeli, G. & Khaliullin, G. Kitaev-Heisenberg model on a honeycomb lattice: possible exotic phases in iridium oxides A 2 IrO 3. *Physical review letters* **105**, 027204 (2010).

50  Catuneanu, A., Rau, J. G., Kim, H.-S. & Kee, H.-Y. Magnetic orders proximal to the Kitaev limit in frustrated triangular systems: Application to Ba3IrTi2O9. *Physical Review B* **92**, 165108 (2015).

51  Becker, M., Hermanns, M., Bauer, B., Garst, M. & Trebst, S. Spin-orbit physics of j= 1 2 Mott insulators on the triangular lattice. *Physical Review B* **91**, 155135 (2015).

52  Li, Y., Winter, S. M., Kaib, D. A., Riedl, K. & Valentí, R. Modified Curie-Weiss law for j eff magnets. *Physical Review B* **103**, L220408 (2021).

53  Rodríguez-Carvajal, J. Recent advances in magnetic structure determination by neutron powder diffraction. *Physica B: Condensed Matter* **192**, 55-69 (1993).

54  Arnold, O. *et al.* Mantid—Data analysis and visualization package for neutron scattering and μ SR experiments. *Nuclear Instruments and Methods in Physics Research Section A: Accelerators, Spectrometers, Detectors and Associated Equipment* **764**, 156-166 (2014).

55  Su(n)ny. https://github.com/SunnySuite/Sunny.jl

56  Dahlbom, D., Miles, C., Zhang, H., Batista, C. D. & Barros, K. Langevin dynamics of generalized spins as su(n) coherent states. *Physical Review B* **106**, 235154 (2022).

57  Dahlbom, D. *et al.* Geometric integration of classical spin dynamics via a mean-field Schrödinger equation. *Physical Review B* **106**, 054423 (2022).

58  Zhang, H. & Batista, C. D. Classical spin dynamics based on SU (N) coherent states. *Physical Review B* **104**, 104409 (2021).

59  Proffen, T. & Welberry, T. Analysis of diffuse scattering via the reverse Monte Carlo technique: A systematic investigation. *Acta Crystallographica Section A: Foundations of Crystallography* **53**, 202-216 (1997).

60  Toth, S. & Lake, B. Linear spin wave theory for single-Q incommensurate magnetic structures. *Journal of Physics: Condensed Matter* **27**, 166002 (2015).

61  Sala, G. *et al.* Field-tuned quantum renormalization of spin dynamics in the honeycomb lattice Heisenberg antiferromagnet YbCl3. *Communications Physics* **6**, 234 (2023).

62  Bordelon, M. M. *et al.* Spin excitations in the frustrated triangular lattice antiferromagnet NaYbO2. *Physical Review B* **101**, 224427 (2020).

63  Jiao, J. *et al.* Quantum Effect on the Ground State of the Triple-Perovskite Ba3MNb2O9 (M= Co, Ni, and Mn) with Triangular-Lattice. *Chemistry of Materials* **34**, 6617-6625 (2022).

64  Xie, T. *et al.* Complete field-induced spectral response of the spin-1/2 triangular-lattice antiferromagnet CsYbSe2. *npj Quantum Materials* **8**, 48 (2023).

65  Scheie, A. O. *et al.* Nonlinear magnons and exchange Hamiltonians of the delafossite proximate quantum spin liquid candidates KYbSe2. *Physical Review B* **109**, 014425 (2024).

66  Rawl, R. *et al.* Ba8CoNb6O24: A spin-1/2 triangular-lattice Heisenberg antiferromagnet in the two-dimensional limit. *Physical Review B* **95**, 060412 (2017).

## Acknowledgements

We acknowledge C. Kim for helpful discussions and S. Martin for helping the implementation of Bayesian optimization algorithm to spin model fitting. This work was supported by the U.S. Department of Energy, Office of Science, Basic Energy Sciences, Materials Sciences and Engineering Division. This research used resources at the Spallation Neutron Source and the High Flux Isotope Reactor, a DOE Office of Science User Facility operated by the Oak Ridge National Laboratory. E.A.G. and C.D.B are supported by the Quantum Science Center (QSC), a National Quantum Information Science Research Center of the U.S. Department of Energy (DOE). L.O.M. and A.E.T. were supported by CONICET under grant no. 3220 (PIP2021). Proof of principle nonlinear spin wave calculations and data collection were supported by the Laboratory Directed Research and Development Program of Oak Ridge National Laboratory, managed by UT-Battelle, LLC, for the U. S. Department of Energy.

## Author contributions

M.B.S, G.B.H, and A.D.C. conceived the project. A.F.M. and J.A.K. synthesized the sample and measured its bulk properties. S.C. conducted the powder neutron diffraction measurement. A.D.C., A.A.P., and M.B.S performed the INS measurement. P.P. analyzed the INS data. P.P. performed LLD and LSWT simulations. J.A.M.P. performed the Onsager reaction field calculation. G.B.H. conducted the NLSWT calculation. E.A.G., C.D.B., A.E.T., and L.O.M. performed the Schwinger boson calculations. All authors participated in the theoretical interpretation and discussion. P.P., M.B.S., G.B.H., and A.D.C. wrote the manuscript with contributions from all authors.

## Competing Interests

The authors declare no competing financial or non-financial interests.

## Tables
No table in the main manuscript.

**Figure captions**

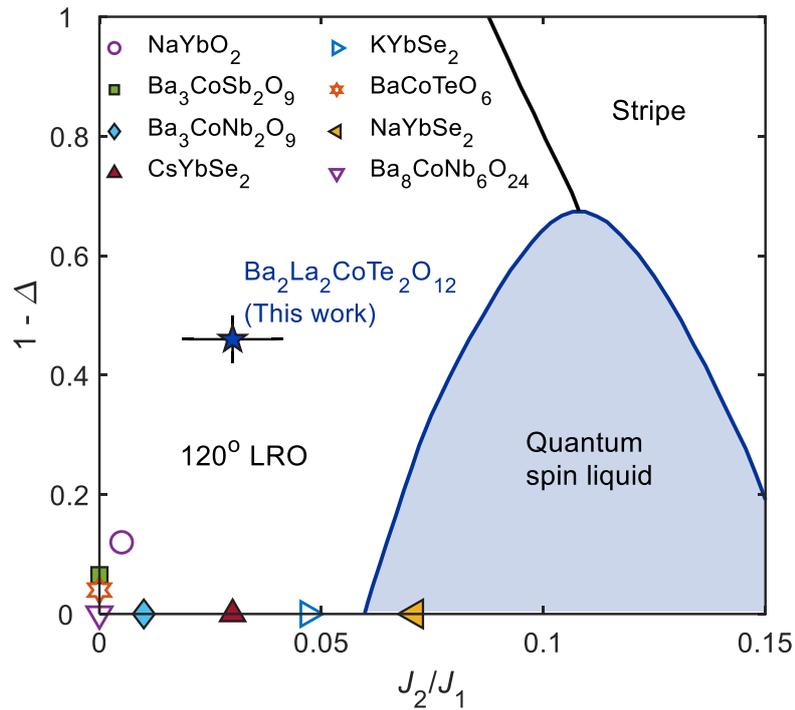

**Fig. 1 | Phase diagram of the $S = 1/2$ XXZ triangular lattice antiferromagnet.** $\Delta = 1$ ($\Delta = 0$) corresponds to the Heisenberg (XY) model. The phase boundaries are obtained from the DMRG calculation in Ref. [19]. The shaded region indicates a quantum spin liquid phase. The two unshaded regions correspond to the 120° and stripe long-range order (LRO), respectively. eight data points excluding $Ba_2La_2CoTe_2O_{12}$ (BLCTO, a solid blue star symbol) are based on the parameters given in previous inelastic neutron scattering studies[15,16,32,62-66]. The error bars associated with BLCTO depict the uncertainties derived from the analysis of our high temperature INS data.

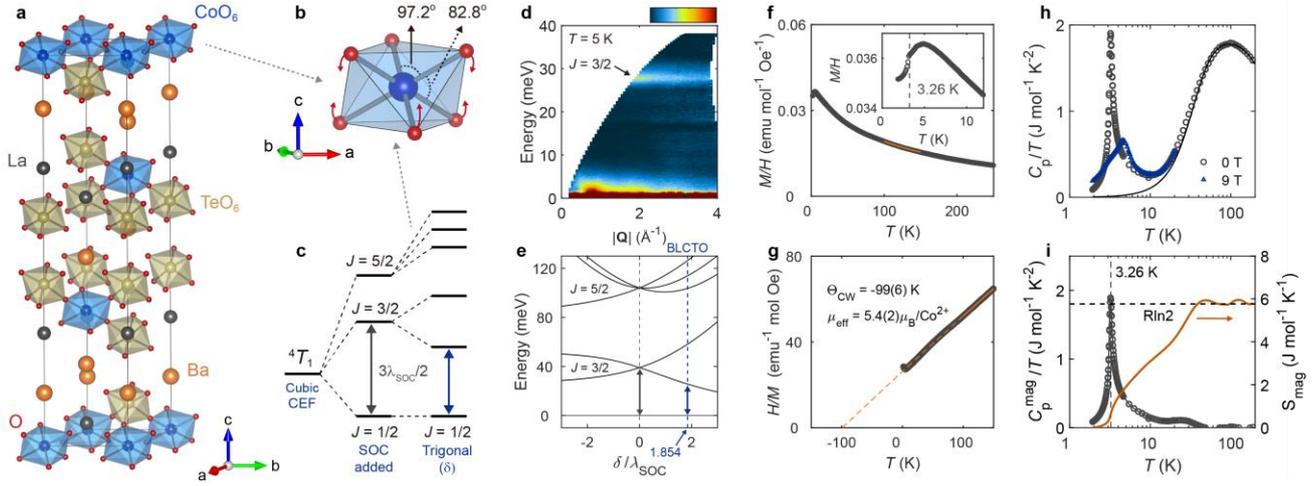

**Fig. 2 | Anisotropic quantum magnetism of $Co^{2+}$ ions in $Ba_2La_2CoTe_2O_{12}$ (BLCTO). a**, Crystallographic unit cell of BLCTO. **b**, Large trigonal compression of a $CoO_6$ octahedron in BLCTO[29]. **c**, Schematic representation of the energy level structure of $Co^{2+}$ with cubic crystal electric field (CEF), spin-orbit coupling (SOC), and trigonal distortion ($\delta > 0$). $\lambda_{soc}$ denotes the single-ion SOC strength of $Co^{2+}$ (~ 26 meV). **d**, The INS spectrum ($I(\mathbf{q},\omega)$, arb. units) of BLCTO measured at 5 K (> $T_N$), demonstrating the presence of $J_{eff} = 1/2 \rightarrow J_{eff} = 3/2$ excitations. **e**, The energy level structure of $Co^{2+}$ with different strength ratios of trigonal distortion ($\delta$) to SOC (see Supplementary Notes). The dashed vertical blue line indicates a value $\delta / \lambda_{SOC} = 1.854$ for BLCTO. **f**, Temperature-dependent magnetization of BLCTO. The inset highlights a phase transition at $T_N = 3.26$ K. **g**, The Curie-Weiss behavior of BLCTO. The solid orange lines in **f** and **g** are the result of curve fitting based on the modified Curie-Weiss law (see Methods). **h**, Temperature-dependent specific heat ($C_p$) of BLCTO under zero (grey circles) and 9 T (blue triangles) magnetic field, presented on a logarithmic scale. The solid black line is the non-magnetic contribution estimated by the Debye model (see Methods). **i**, Remnant specific heat after subtraction of the non-magnetic component from **h** (grey circles) and the corresponding magnetic entropy (solid orange line).

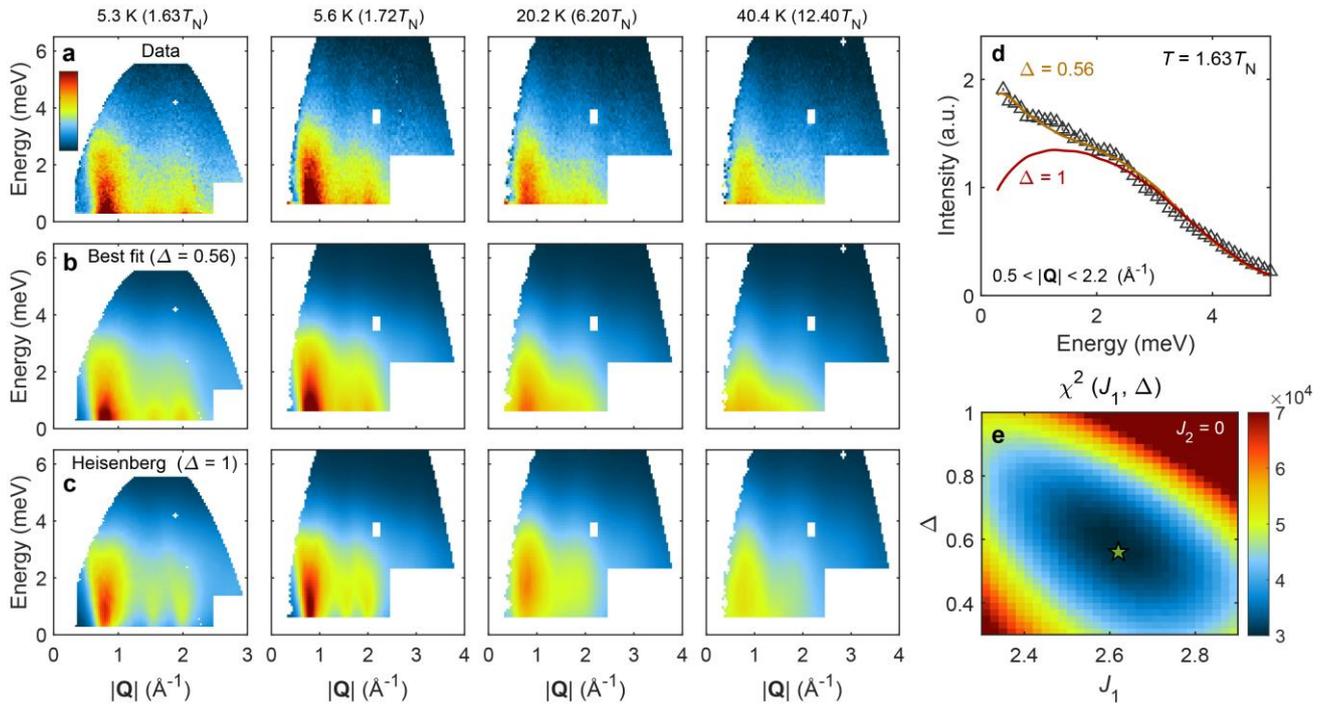

**Fig. 3 | Quantitative analysis of magnetic excitations in the paramagnetic phase of BLCTO ($T > T_N$). a**, The INS data for BLCTO ($I(|q|,\omega)$, arb. units) measured at various temperatures above $T_N$. The incident neutron energies of the four datasets were 6, 10, 10, and 10 meV, respectively. **b**, Optimal fit of the LLD simulation to the data in **a**, resulting from a parameter set $J_1 = 2.62$ meV, and $\Delta = 0.56$. **c**, Excitation spectra calculated from the LLD simulation with $J_1 = 2.62$ meV, and $\Delta = 1$ (Heisenberg). **d**, Comparison between the measured energy-dependent paramagnetic scattering profile and the LLD simulations for $\Delta = 0.56$ and 1. Standard deviations of the data points are much smaller than the data symbol. **e**, The goodness-of-fit quantified through chi-square ($\chi^2$) for different $J_1$ and $\Delta$ values (see Methods). The position of the minimal $\chi^2$ ($J_1 = 2.62$ meV & $\Delta = 0.56(2)$) is indicated by a green star.

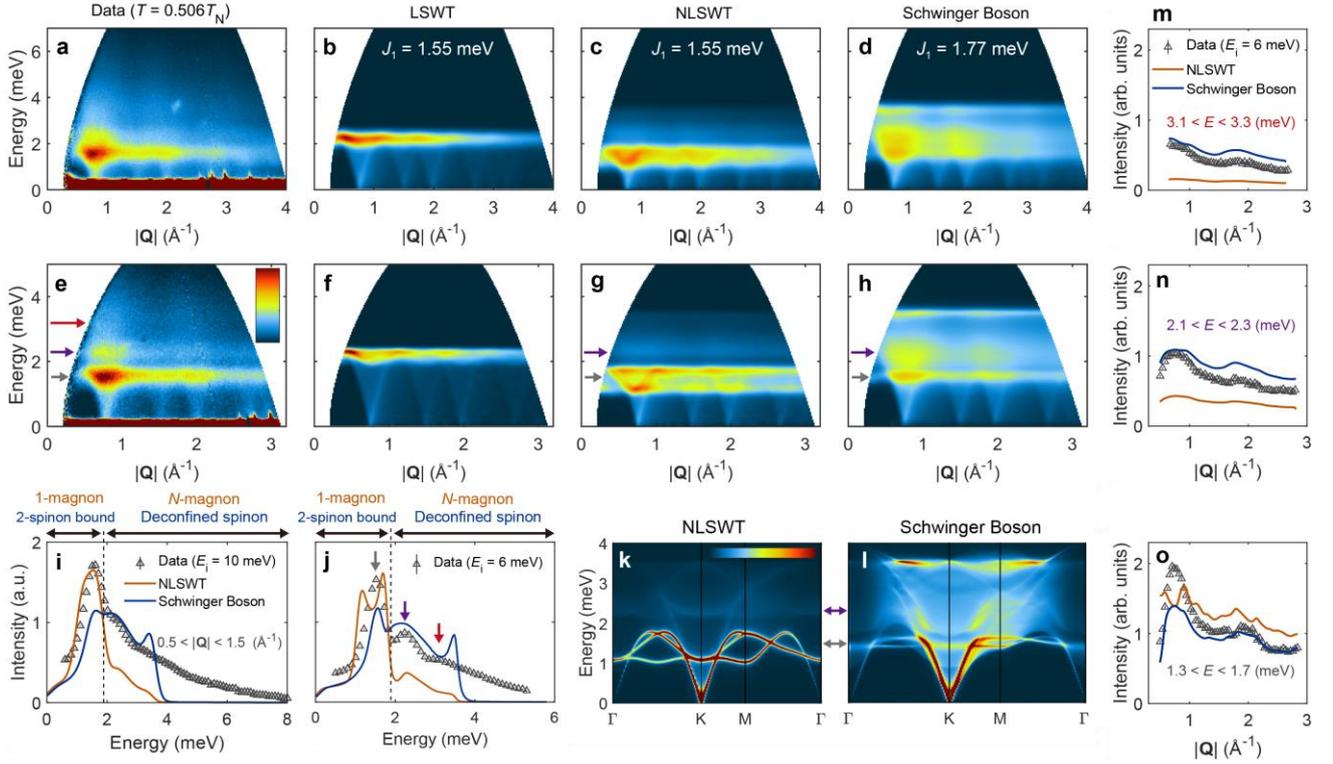

**Fig. 4 | Magnetic excitations of BLCTO in the 120° ordered phase ($T < T_N$). a**, INS data ($I(|\mathbf{q}|,\omega)$, arb. units) measured at 1.65 K (= $0.506 T_N$) with $E_i$ = 10 meV. **b–d**, Corresponding INS cross-sections calculated from linear spin-wave theory (LSWT), non-linear spin-wave theory (NLSWT), and Schwinger boson theory. The calculations were conducted with $J_1$ = 1.55 meV or 1.77 meV, $\Delta$ = 0.56, and $J_2$ = 0. **e**, INS data measured at 1.65 K (= $0.506 T_N$) with $E_i$ = 6 meV. **f–h**, Equivalent to **b–d**, but convoluted with the coverage and instrumental resolution of the data in **e**. **i–j**, Measured and calculated energy-dependent INS cross-sections in the low-$|\mathbf{Q}|$ region (0.5 Å$^{-1}$ < $|\mathbf{Q}|$ < 1.5 Å$^{-1}$). The dashed vertical lines in **i** and **j** denote the one-magnon bandwidth. **k–l**, Single-crystal excitation spectra along high-symmetry directions ($I(\mathbf{q},\omega)$, arb. units), calculated by NLSWT and Schwinger boson theory, respectively. **m–o**, $|\mathbf{Q}|$ dependence of the measured ($E_i$ = 6 meV) and calculated INS cross-sections at the energies indicated by the red, purple, and grey arrows in **e** and **j**. Standard deviations of the data points in **i–j** and **m–o** are much smaller than the data symbol. Only the calculation results in panels **a–j** and **m–o** are convoluted with instrumental resolution.

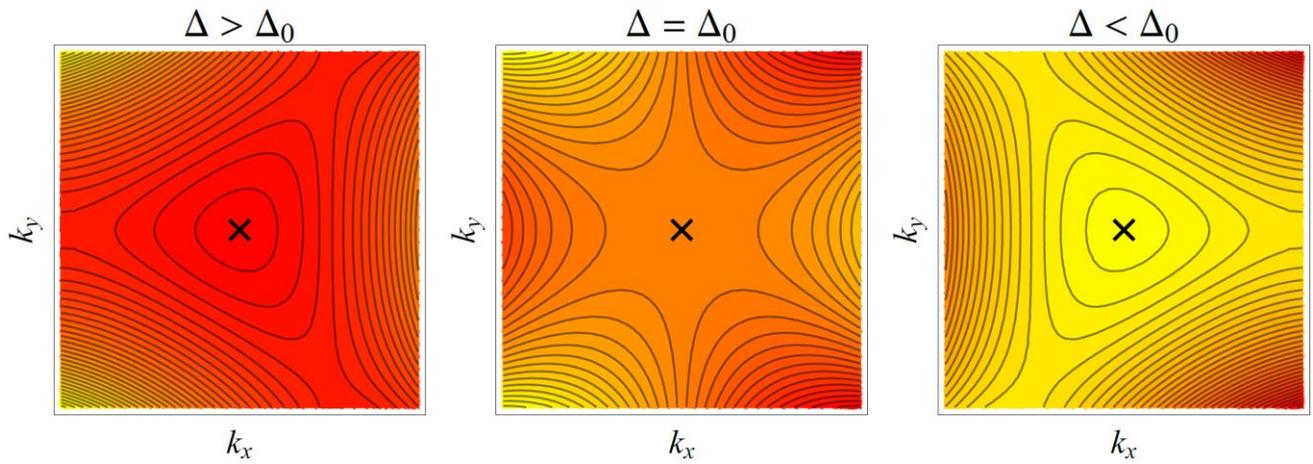

**Fig. 5 | Flat magnon dispersion ("monkey saddle") around the K points induced by planar XXZ anisotropy ($\Delta$).** Magnon energy $E$ against magnon momentum ($k_x$, $k_y$) around one of the K points (black cross) for $\Delta > \Delta_0$ (left), $\Delta = \Delta_0$ (center), and $\Delta < \Delta_0$ (right), as given by Eq. (2) for $c_2 > 0$, $c_2 = 0$, and $c_2 < 0$, respectively. While the K point is a local minimum (maximum) for $\Delta > \Delta_0$ ($\Delta < \Delta_0$), it realizes a higher-order Van Hove singularity at the critical anisotropy $\Delta = \Delta_0$. Note that smaller (larger) magnon energies are marked by red (yellow) color.